\documentclass [preprint2]{aastex}

\newcommand{\cbstart}{\relax}
\newcommand{\cbend}{\relax}
\newcommand{\CapOne} {We show the HETGS flux spectrum of II~Pegasi.
  The spectrum is the combined HEG and MEG, diffracted orders -3 to
  +3. It has been smoothed for presentation. Some prominent lines have
  been labeled.  }

\newcommand{\CapTwo} {The top panel shows the full-band count-rate
  light curve. The smooth curve is a two-ribbon Solar flare model.
  The bottom shows the contnuum light curve in a narrow,
  short-wavelength bandpass.  Symbols show the light curve in net line
  fluxes for a few features.  Note the decreased amplitude of the
  lines, relative to the continuum.
  }

\newcommand{\CapThree} {We show the model of the differential emission
  measure for the flare, quiescent, and total spectrum.  Differences
  in the fit from $\log T=6.5-7.2$ are indicative of uncertainties in
  the fitting.  The large, hot peak is the flare, which did not affect
  the cooler region of the DEM.  Overplotted points, connected with
  broken lines, are the DEM determinations of \citet{GRIFF98} and
  \citet{Mewe1997}. 
  }

\newcommand{\CapFour} {We show the modulation of line and continuum
  between flare and quiescent states.  The modulation is defined as
  $[f(flare)-f(quiescent)]/[f(flare)+f(quiescent)]$. It is 1.0 for
  $f(quiescent)=0$, and 0.0 for $f(quiescent)=f(flare)$.  Lines have
  been grouped by $\log T_{\mathrm{max}}$, and clearly show increasing
  modulation with temperature, meaning that the flare was very hot.
  Features which give a better than 2.5$\sigma$ confidence in the
  modulation have been emphasized.
  
  The continuum modulation is defined similarly, except that
  $T_{\mathrm{max}}$ is not well defined.  Instead, we use a
  pseudo-temperature derived from the continuum band wavelength ($T =
  hc/k\lambda$), primarily to enable placement of continuum data on
  the same axis.  The continuum is strongly modulated with about equal
  amplitude in each wavelength band.  It is not very sensitive to
  temperature because its emissivity distribution is broad --- a hot
  continuum adds significant flux at all wavelengths above a lower limit
  (or, at all lower pseudo-temperatures).
  }

\newcommand{\CapFive} {
  Panels a-d show detailed comparison of the observed
  counts spectrum and the model folded through the instrumental
  response. This is the spectrum for the entire observation and uses
  the fitted DEM and abundances.  The spectra have been
  smoothed by a Gaussian kernel with width equal to the instrument
  resolution (0.008\AA\ and 0.004\AA\ Gaussian $\sigma$'s for MEG and
  HEG, respectively). The spectra were binned to 0.005 \AA\ (MEG) and
  0.0025 \AA\ (HEG). The upper curves are MEG counts, and the lower
  HEG. The dotted curve is the model.  Up to eight of the brightest
  features in the APED emissivity tables for the given model have been
  labeled in each graph.
  }
\newcommand{\CapFiveB} {See 5a.}
\newcommand{\CapFiveC} {See 5a.}
\newcommand{\CapFiveD} {See 5a.}

\newcommand{\Note}[1]{
  \advance\textwidth-72pt\par
  \parbox{\textwidth}{
    \hfill\Putline
    {\small\sf#1\par}
    \Putline\par
    }
  \advance\textwidth+72pt\par
}
\newcommand{\Putline}{\rule{\the\textwidth}{1pt}  }

\newcommand{\iip}{II~Peg}

\shortauthors{Huenemoerder, et al.}
\shorttitle{X-Ray Spectroscopy of II~Pegasi}

\begin{document}

\title{X-Ray Spectroscopy of II Pegasi: Coronal Temperature Structure,
  Abundances, and Variability} 

\author{
  David~P.~Huenemoerder\altaffilmark{1}\altaffiltext{1}{dph@space.mit.edu},
  Claude~R.~Canizares\altaffilmark{2}\altaffiltext{2}{crc@space.mit.edu}, and
  Norbert~S.~Schulz\altaffilmark{3}\altaffiltext{3}{nss@space.mit.edu}
  }
\affil{MIT Center for Space Research\\
70 Vassar St., \\
Cambridge, MA  02139}

\begin{abstract}
  
  We have obtained high resolution X-ray spectra of the coronally
  active binary, II~Pegasi (HD~224085), covering the wavelength range
  of 1.5-25\AA.  For the first half of our 44 ksec observation, the
  source was in a quiescent state with constant X-ray flux, after
  which it flared, reaching twice the quiescent flux in 12 ksec, then
  decreasing.  We analyze the emission-line spectrum and continuum during
  quiescent and flaring states.  The differential emission measure
  derived from lines fluxes shows a hot corona with a continuous
  distribution in temperature.  During the non-flare state, the
  distribution peaks near $\log T = 7.2$, and when flaring, near
  $7.6$.  High-temperature lines are enhanced slightly during the
  flare, but most of the change occurs in the continuum.  Coronal
  abundance anomalies are apparent, with iron very deficient relative
  to oxygen and significantly weaker than expected from photospheric
  measurements, while neon is enhanced relative to oxygen.  We find no
  evidence of appreciable resonant scattering optical depth in line
  ratios of iron and oxygen.  The flare light curve is consistent with
  Solar two-ribbon flare models, but with a very long reconnection
  time-constant of about 65 ks.  We infer loop lengths of about 0.05
  stellar radii, to about 0.25 in the flare, if the flare emission
  originated from a single, low-density loop.

\end{abstract}

\keywords{
stars: coronae --- stars: individual (II~Pegasi) --- X-rays: stars ---
stars: abundances --- stars: activity --- line: identification
}

%\twocolumn

\section{Introduction}

II~Pegasi (HD~224085) is a 7.6 $V$ magnitude spectroscopic binary
comprised of a K2-3~V-IV star plus an unseen companion in a 6.7 day
orbit \citep{CABS93,Berdy98}.  II~Peg is bright and active in the
radio, optical, UV, and X-ray regions \citep{Owen78, Walter81,
  Schwartz81}.  It was discovered as photometrically variable by
\citet{Chugainov76}, and was classified as an RS~CVn system by
\citet{Rucinski77} based on the photometric properties and by nature
of its strong optical emission lines.

It has been known for quite some time that the RS~CVn class is very luminous
in X-rays and that this luminosity is strongly correlated with stellar
rotation \citep{Walter81}. The working paradigm is that the photometric
and spectroscopic features are due to scaled-up versions of Solar type
``activity'': dark photospheric spots, bright chromospheric plages and
prominences, a UV-bright transition region, an X-ray-bright corona,
and variability from short-lived flares to long-term cycles.

One of the fundamental outstanding problems of stellar activity is the
nature of the underlying coronal heating mechanisms, which are
ultimately tied to the stellar rotation and a magnetic dynamo.  We
cannot observe these directly but infer their presence through
observations of the energy emitted, correlations with fundamental
stellar parameters, and through time variability.  The {\em Chandra}
High Energy Grating Spectrometer (HETGS) provides new capabilities in
X-ray spectroscopy, giving us much enhanced spectral resolving power
and sensitivity than previous X-ray observatories. Its performance is
complementary to other current instruments such as the {\em Chandra}
Low Energy Grating Spectrometer and the {\em XMM} Reflection Grating
Spectrometer.

The Chandra HETGS spectra of \iip\ provide new and definitive
information in several key areas pertinent to the hottest part of the
outer stellar atmosphere, the corona:  the coronal iron abundance is
low; the differential emission measure is continuous in temperature;
flare time profiles are consistent with Solar two-ribbon flare models
and are very hot.  Here we present the HETGS spectrum, line flux
measurements, differential emission measure and abundance fits,
density diagnostics, a model spectrum, and line and continuum
light-curves.

\section{Observations and Data Processing}
\label{sec:obsdp}

\subsection{The Chandra HETGS}
\label{sec:obshetgs}

The HETG assembly \citep{Markert94, Canizares01} is comprised of an
array of grating facets that can be placed into the converging X-ray
beam just behind the Chandra High Resolution Mirror Assembly (HRMA).
When in place, the gratings disperse the X-rays creating spectra that
are recorded at the focal plane by the spectroscopic array, ``S'', of
ACIS (Advanced CCD Imaging Spectrometer).  There are two different
grating types, designated MEG and HEG, optimized for medium and high
energies, respectively, which overlap in spectral coverage.  The HETGS
provides spectral resolving power of $\lambda/\Delta\lambda
=100-1000$. The line width is about 0.02 \AA\ for MEG, and 0.01 \AA\ 
for HEG (full widths, half maximum).  The effective area is ~1-180
$\mathrm{cm^2}$ over the wavelength range of 1.2-25~\AA\ (0.4-10 keV).
Multiple overlapping orders are separated using the moderate energy
resolution of the ACIS detector.  (For detailed descriptions of the
instrument we refer you to the ``Proposers' Observatory
Guide,''\footnote{Available from {\tt
    http://cxc.harvard.edu/udocs/docs/docs.html}}).

\subsection{The Observation}
\label{sec:obsobs}

\iip\ was observed on 17-18 October, 1999 (observation identifier
1451, Sequence Number 270401).  At the time of the observation, the
ACIS camera had a problem with one of its Front End Processors (FEP).
This required the omission of one of the six CCDs of ACIS-S from
telemetry.  The S0 chip, at the longest wavelength on the negative
order side was switched off, with a corresponding loss of some spectral
coverage not redundant with the plus side and some effective area
where plus and minus overlap.  The observation was otherwise done in
nominal, timed-exposure mode.

The count rate in the dispersed order region is about $3\ 
\mathrm{counts\,s^{-1}}$, averaged over the length of good exposure of
44,933 seconds.  The zeroth-order is to ``piled'' to be useful.
(``Pileup'' is a term referring to the unresolved temporal coincidence
of photons in a $3\times3$ pixel cell, which confounds the accurate
determination of the photon energy or gives the wrong pixel pattern.
See \citet{Davis2001} for detailed definitions and modeling
techniques.)

We use the ephemeris of \citet{Berdy98}: the epoch (which defines
phase$=0.0$) is HJD~2449582.9268, and the period is 6.724333 days.
Our observed range of phases is 0.56--0.64.  An optical light curve
one year later \citep{Tas2000} shows the brightness increasing and
about mid-way between the maximum and minimum at these phases, with
with largest optical brightness modulation ever observed for II~Peg.
We have not yet seen an optical light-curve contemporaneous with the
X-ray observations.

The source varied strongly during the observation. About half-way
through the exposure, a flare occurred (see
Section~\ref{sec:alightcurve}).  We separately analyze the pre-flare
interval (first 22 ks), the flaring interval (last 17 ks), and also
the entire observation.

\subsection{Data Processing}
\label{sec:obsdataproc}

Event lists were processed with several versions of the CIAO (Chandra
Interactive Analysis of Observations) software suite.  The pipeline
standard processing used version {\tt CIAO 1.1}.  Custom re-processing
was done with various versions of the CIAO tools and the calibration
database (CALDB), which are ultimately equivalent to {\tt CIAO 2.0}
and {\tt CALDB 2.0}.  Further analysis products were made with the
CIAO tools to filter bad events (based on grade and bad pixels), bin
spectra, make light-curves, make two-dimensional images in wavelength
and time, and to make responses.

\section{Analysis}
\label{sec:analysis}

Analysis of products produced with CIAO tools was performed either with
ISIS\footnote{ISIS is available from {\tt
    http://space.mit.edu/CXC/ISIS}} \citep{ISIS2000},
which was specifically designed for use on Chandra grating data and as
an interface to the APED (``Astrophysical Plasma Emission Database'';
\citet{Smith98}), or with custom programs written in {\em IDL}
(``Interactive Data Language'') using the ISIS measurements and APED
emissivities.  We used APED version 1.01 in conjunction
with the Solar abundances values of \citet{Anders89} and the ionization
balance of \citet{MM98}.

\subsection{Line Fluxes}
\label{sec:alineflux}

We measured line fluxes in ISIS by fitting emission lines with a
polynomial plus delta-function model convolved by the instrument
response.  Plus and minus first orders from HEG and MEG were fit
simultaneously.  Regions of very low counts were grouped by two to
four bins from the standard-product scales of 0.005 (MEG) or 0.0025
(HEG) \AA/bin.  The free parameters were the line wavelengths, the
continuum level, and the line fluxes.  Up to three line components
were fit in some blends or groups.  The background is low enough to be
neglected, being much less than a count per bin per 50 ks.
Statistical 68\% confidence limits are computed for each free
parameter and are converted into equivalent Gaussian dispersions,
since they are predominantly symmetric.

The grating response matrices are a convenient encoding of the
instrumental line-profile as a function of wavelength.  The response
was represented by single-Gaussian profiles.  A poorer fit in the
wings is apparent in the strongest lines, in is most likely due to
inaccuracies in determination and parameterization of the true profile.
This does not significantly affect the flux measurements.

The overall effective area calibration is generally accurate to better
than 10\%, but has some systematic uncertainties between the different
ACIS-S CCDs amounting to about 20\% in some regions.  The absolute
wavelength scale calibration is accurate to about 50-100
$\mathrm{km\,s^{-1}}$, which allows unambiguous line identification.

\begin{figure}
   \figurenum{1}
  \epsscale{1.0}
%  \plotone{f1.eps}
 % \plotone{Full_spec.eps}
 % \plotone{Full_spec.pdf}
$$ \includegraphics[scale=1.2,angle=-90]{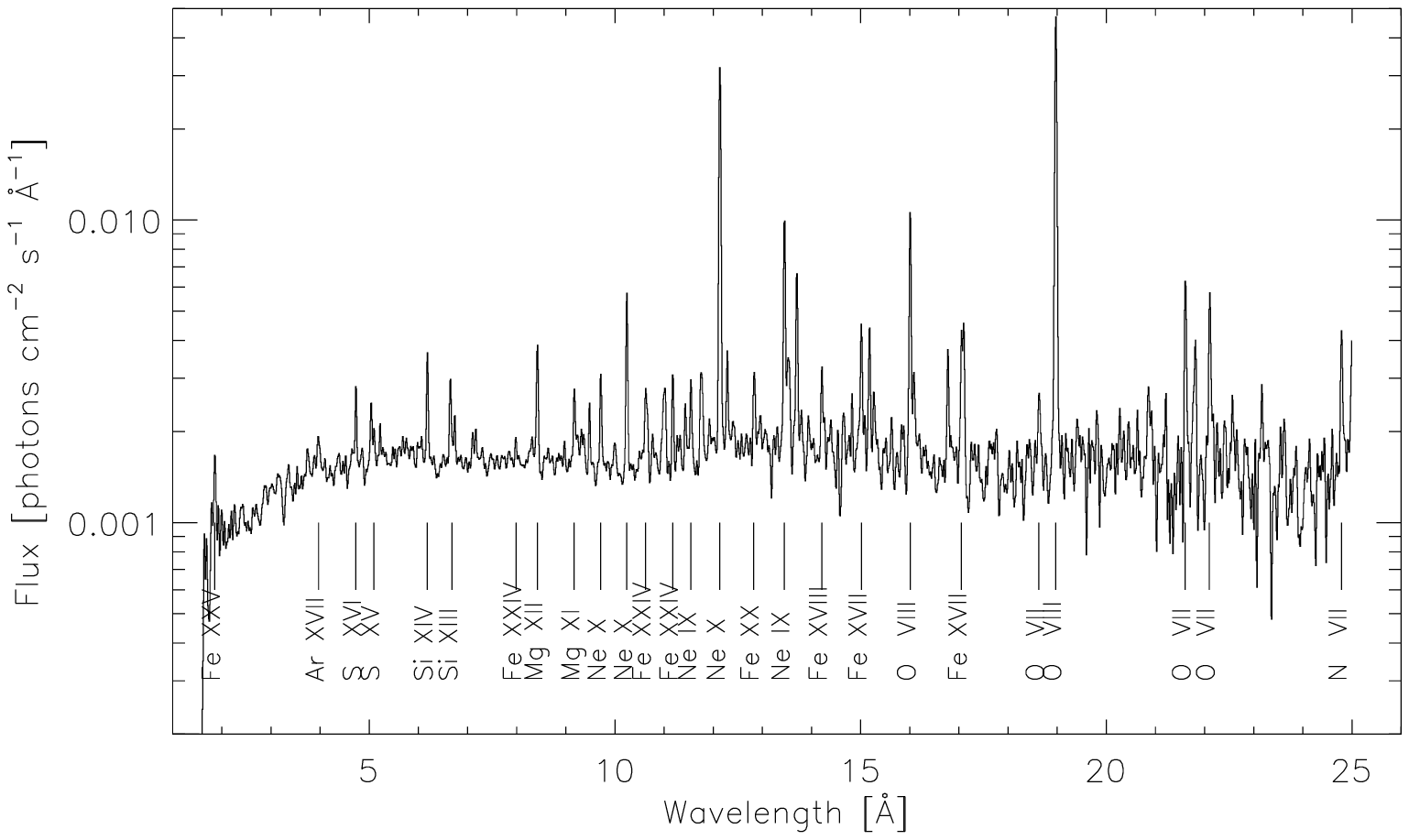}$$
  \caption{\CapOne}
  \label{fig:fullspec}
  \end{figure}

We list the line fluxes, statistical uncertainties, and identifications
in Table~\ref{tab:linelist}.  The full spectrum is shown in
Figure~\ref{fig:fullspec}.

\subsection{Differential Emission Measure and Abundances}
\label{sec:adem}

The flux which we determine is an integral over the emitting plasma
volume along the line of sight.  The plasma could have a range of
temperatures, densities, velocities, geometric structures, or
non-equilibrium states.  We assume it is in Collisional Ionization
Equilibrium (CIE). 
                                \cbstart % refreply begin 1 [2001.05.09]
This is not synonymous with thermodynamic equilibrium in which every
process is in detailed balance with its inverse. It instead refers to
a stable ionization state of a thermal plasma under the {\em coronal
  approximation}, in which the dominant processes are collisional
excitation and ionization from the ground state, and radiative and
dielectronic recombination.  In other words, photoexcitation and
photoionization are not important, nor are collisional excitation or
ionization from excited states. (See \citet{McCray1987} for a good
tutorial on thermal plasma processes.)  The latter assumption is not
strictly true, since there are some lines which are density sensitive
through metastable levels, most notably, the helium-like forbidden and
intersystem transitions.  We treat these separately.

We expect that CIE holds through most of the observation, even though
a moderate size flare occurred. If the plasma can ionize or recombine
quickly enough relative to an integration time, then we will see some
average steady state ionization. Some characteristic times for line
appearance through ionization or recombination are given by
\citet{Golub1989} in their Table~V (columns $\tau_\mathrm{app}$ and
$\tau_\mathrm{recom}$). Most features respond in seconds to hundreds of
seconds. The main exception is Mg~{\sc x} in the extended corona and
polar plumes, which are low density structures. \citet{Mewe1985} and
\citet{Doschek1980} also discuss transient ionization affects in Solar
flares.

In our analysis, we did split the spectrum into pre-flare and flare
states, omitting the time of rapidly increasing flux to provide some
insurance against assumption of inappropriate models during a heating
phase, in addition to measurement on the spectrum integrated over the
entire observation.

We stress, however, that CIE is primarily an assumption.  Conditions
on stars may not be similar to the Sun in density and energy input
profiles. We will assume CIE until we can demonstrate that it fails as
a model.
                                \cbend % refreply end 1 [2001.05.09]

The emitted flux can be expressed as 
                                \cbstart % refreply begin 1 [2001.05.09]
({\em cf.} equations 1 and 3 in \citet{GRIFF98})
                                \cbend % refreply end 1 [2001.05.09]
 \begin{equation}
   \label{eq:vembasic}
   f_l = \frac{A_l}{4\pi d^2}\,\int{dV\,G_l(T_e,n_e) n_e n_H}.
 \end{equation}
 The flux in feature $l$ is $f_l$, $A_l$ is the elemental abundance,
 $d$ is distance, $n_e,\, n_h$ are electron and hydrogen densities,
 $V$ the volume, $G_l$, is the feature emissivity (in units of
 $\mathrm{photons\,cm^{3}\,s^{-1}}$), and is a function of electron
 temperature, $T_e$, and electron density.  This formulation also
 requires that the coronal approximation be valid: that the plasma is
 optically thin and in collisional equilibrium.
 The function $G$ contains all the fundamental atomic data as well as
 an ionization balance.  This expression is often approximated by
 using a mean value of $G$, since it is typically sharply peaked over
 a small temperature range. We can write
\begin{equation}
  \label{eq:vemapprox}
%% %-obsolete
%  f_l =  \frac{A_l}{4\pi d^2}\, k(\Delta T)\,\overline{G}(T_\mathrm{max})\, E(T_\mathrm{max})
%% %+obsolete
  f_l =  \frac{A_l}{4\pi d^2}\, \overline{G}_l\, E(T_\mathrm{max}).
\end{equation}
$\overline{G}$ is a mean emissivity over the temperature range of
maximum emissivity, and $E$ is defined as the ``volume emission
measure'', or VEM at the temperature of maximum emissivity,
$T_\mathrm{max}$.  Dependence on $n_e$ has been ignored; many emission
lines are only very weakly dependent upon $n_e$ in the expected
coronal temperature and density regimes.

It is often more useful to transform the emission measure to a
differential form in $\log T$, by restating the volume element, $dV$,
as  a function of temperature, $dV(T)$. We then substitute $dV =
(dV/d\log T)\,d\log T$ to derive the form
 \begin{equation}
   \label{eq:dembasic}
   f_l = \frac{A_l}{4\pi d^2}\,
   \int{d\log T\,G_l(T) \left[n_e n_h \frac{dV}{d\log T}\right]}.
 \end{equation}
 The quantity in square brackets is the ``differential emission
 measure'' (DEM), which we denote by $D(T)$.
 
 The emission measure is a fundamental quantity of the plasma to be
 determined, since it represents the the underlying balance between
 the input heating and the energy losses.  The integral relation for
 $D$ cannot be inverted uniquely, given the physical form of $G$.  The
 functions, $G_l(T)$, are effectively a set of basis functions, but
 they largely overlap and they are not orthogonal.  The mathematics of
 this equation in the astrophysical context has been studied by
 \citet{CraigBrown76}, \citet{HubenyJudge95}, and \citet{McIntosh98}.
 Inversion is possible, but regularization is required to obtain a
 meaningful solution. 
%% %+obsolete
% More often, the approximate method of
%  equation~\ref{eq:vemapprox} is applied iteratively
%  \citep[e.g.]{GRIFF98}, or even a simpler isothermal, upper-limit
%  solution.  At the other extreme of complexity, \citet{Kashyap98} used
%  Baysian techniques and a Monte-Carlo Markov-Chain solution.
 
%  Each technique has its advantages and disadvantages.  The isothermal
%  upper-limit is quick, but it can suppress fairly large amplitude
%  structure in the DEM.  Bayesian techniques are becoming popular since
%  they promise to automatically include prior information and provide
%  parameter uncertainties, but they have a large overhead since the
%  marginal distributions of each model parameter are required, and it
%  is not often clear that the advantages balance the effort \citep[see
%  the discussion in][section 10.1]{Gersh99}.
%% %-obsolete
 
 We have adopted a simple $\chi^2$ technique to determine the DEM, and
 we minimize a discrete form of Equation~\ref{eq:dembasic}:
 \begin{equation}
   \label{eq:demchisq}
   \chi^2 = \sum_{l=1}^{L}
             \frac{1}{\sigma_l^2}
             \left[f_l - A_{Z(l)}\frac{\Delta\log T}{4\pi d^2}
             \sum_{t=1}^{N}{G_{lt}\Psi(e^{\ln D_t},k)} \right]^2
 \end{equation}
 Here, $l$ is a spectral feature index, and $t$ is the temperature
 index. The measured quantities are the $f_l$, with uncertainties
 $\sigma_l$. The {\em a priori} given information are the
 emissivities, $G_{lt}$, and the source distance, $d$.  The
 minimization provides a solution for $D_t$ and $A_Z$.  The
 exponentiation of $\ln D$ forces $D_t$ to be non-negative, and
 $\Psi$ is a smoothing operator.
%% %+obsolete
%  which imposes a prior
%  bias that the resulting DEM should not have sharp curvature, and also
%  serves to reduce the formal degrees of freedom, which can be
%  arbitrarily set by the number of temperature grid points regardless
%  of actual number of independent spectral features.  
%% %-obsolete
 $A_Z$ is the
 abundance of element $Z$. We omit
 spectral features which are line blends of different
 elements and of comparable strengths.  We use a temperature grid of
 60 points spaced by 0.05 in $\log T$, from $\log T=5.5$ to 8.5.  
%% %+obsolete
% This
%  over-samples the APED grid by about a factor of two, but produces a
%  better behaved solution with regard to edge effects and smoothing.
%
%  Our smoothing function is a convolution by a Gaussian kernel.  We
%  find that a kernel full-width, half-maximum of five bins behaved
%  well.  This corresponds to a $\Delta\log T$ of 0.25, which is the
%  width of the emissivity curves at about 80\% of their maximum.  To
%  minimize $\chi^2$, we use the {\tt AMOEBA} function in {\em IDL},
%  which is based on the routine of the same name from \citet{Press86},
%  and uses the downhill simplex method of \citet{Nelder65}.  The
%  amoeba-simplex optimization is relatively slow compared to other
%  optimization methods, but it does not require numerical derivatives.
%% %-obsolete
 For the initial values of the DEM, we applied
 Equation~\ref{eq:vemapprox}.
 
 In order to get convergent fits, we need to include as many lines as
 possible to span a broad temperature range.  The fit is not
 constrained where emissivities go to zero, so filling in the gaps
 with any available lines is crucial.  For this reason, the relative
 abundances must be included as model parameters.  The solution
 determines relative abundances and a temperature distribution, but
 has an arbitrary normalization.  We remove this degeneracy
 by examining line-to-continuum ratios. We scale the DEM so that
 the model continuum agrees with the data, and then scale the
 abundances inversely. We use a strong feature near the maximum
 sensitivity, specifically, Mg~{\sc xi} at 9.2\AA.
 
 We assess uncertainties in the fit by running simulations with
 different trial DEM distributions, from flat to sharply peaked.  The
 input DEM is used to generate line fluxes to which noise is added,
 and which were then re-fit. We find that peaks in the trial DEM are
 determined to 0.1 in $\log T$ of the true position, and the fitted
 distribution does follow the model, with a mean deviation of about
 20\%. Single sharp peaks are reproduced, with a minimum width (FWHM)
 of about 0.15 in $\log T$.  We also explored variations in the input
 abundances of our trial models.  The fitted abundances reproduce the
 input values to an accuracy of 20\%.

\subsection{Light Curves}
\label{sec:alightcurve}

\begin{figure}
   \figurenum{2}
  \epsscale{1.0}
%  \plotone{f2.eps}
  \plotone{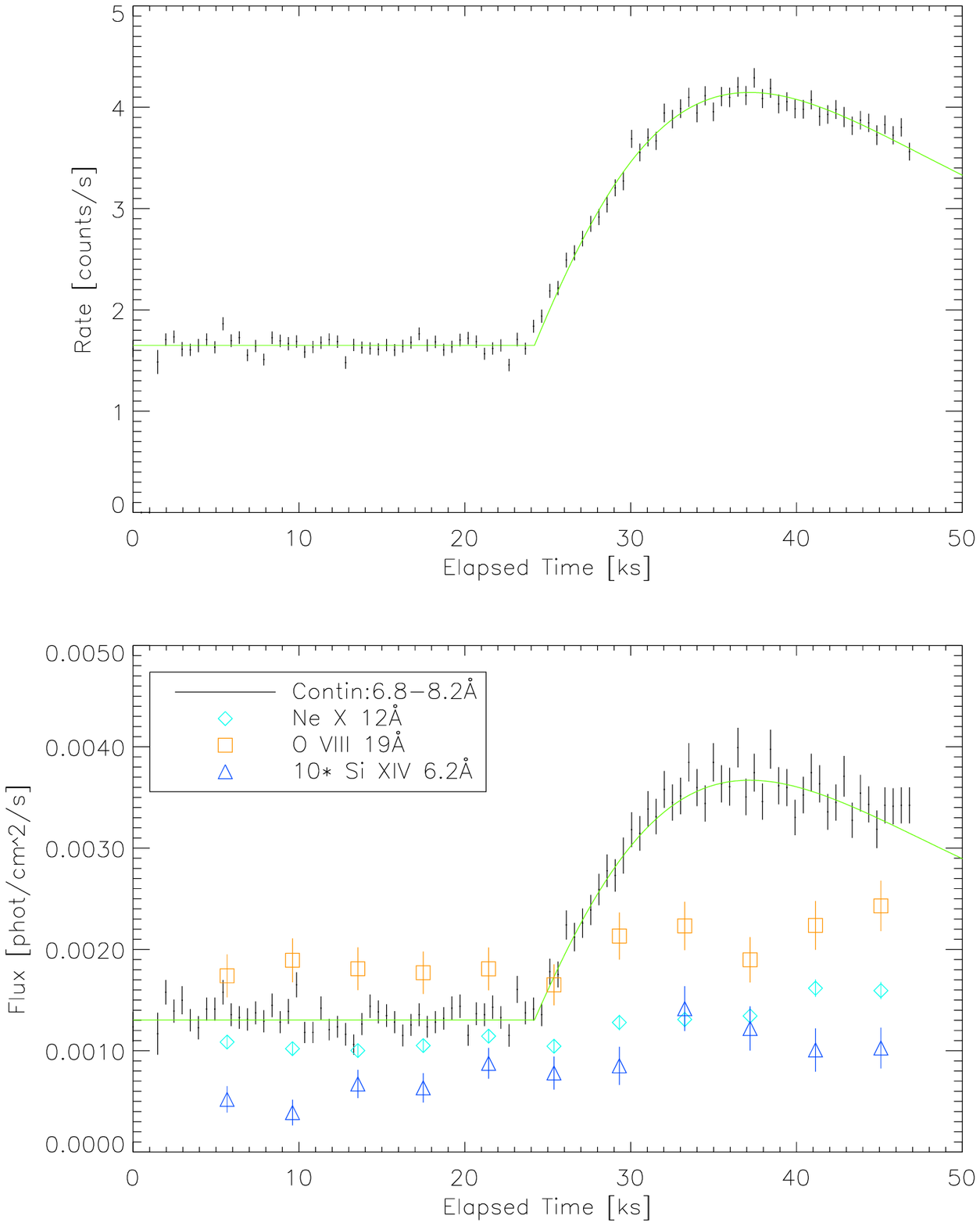}
  %\plotone{Lc_basic.eps}
 % \plotone{Lc_basic.pdf}
  \caption{\CapTwo}
  \label{fig:lcbasic}
  \end{figure}

We made light curves in line and continuum band-passes by binning
events from the default extraction regions, including diffracted
orders $-3$ to $+3$ (first orders contain all but a few percent of the
counts) and both HEG and MEG spectra.  To convert rates to fluxes, we
used mean effective areas for the entire observation, since our time
bins are large with respect to the dither period (lines bands), or
band-passes are large relative to non-uniformities (continuum bands).
Figure~\ref{fig:lcbasic} shows the count-rate integrated over the
entire bandpass (1--25 \AA), and the flux for a narrower continuum
band and for selected lines.  Line flux curves are net flux, less a
local continuum rate scaled to the line's bandpass.

\section{Results}

\subsection{Temperature Structure}

\begin{figure}
   \figurenum{3}
  \epsscale{1.0}
%  \plotone{f3.eps}
  \plotone{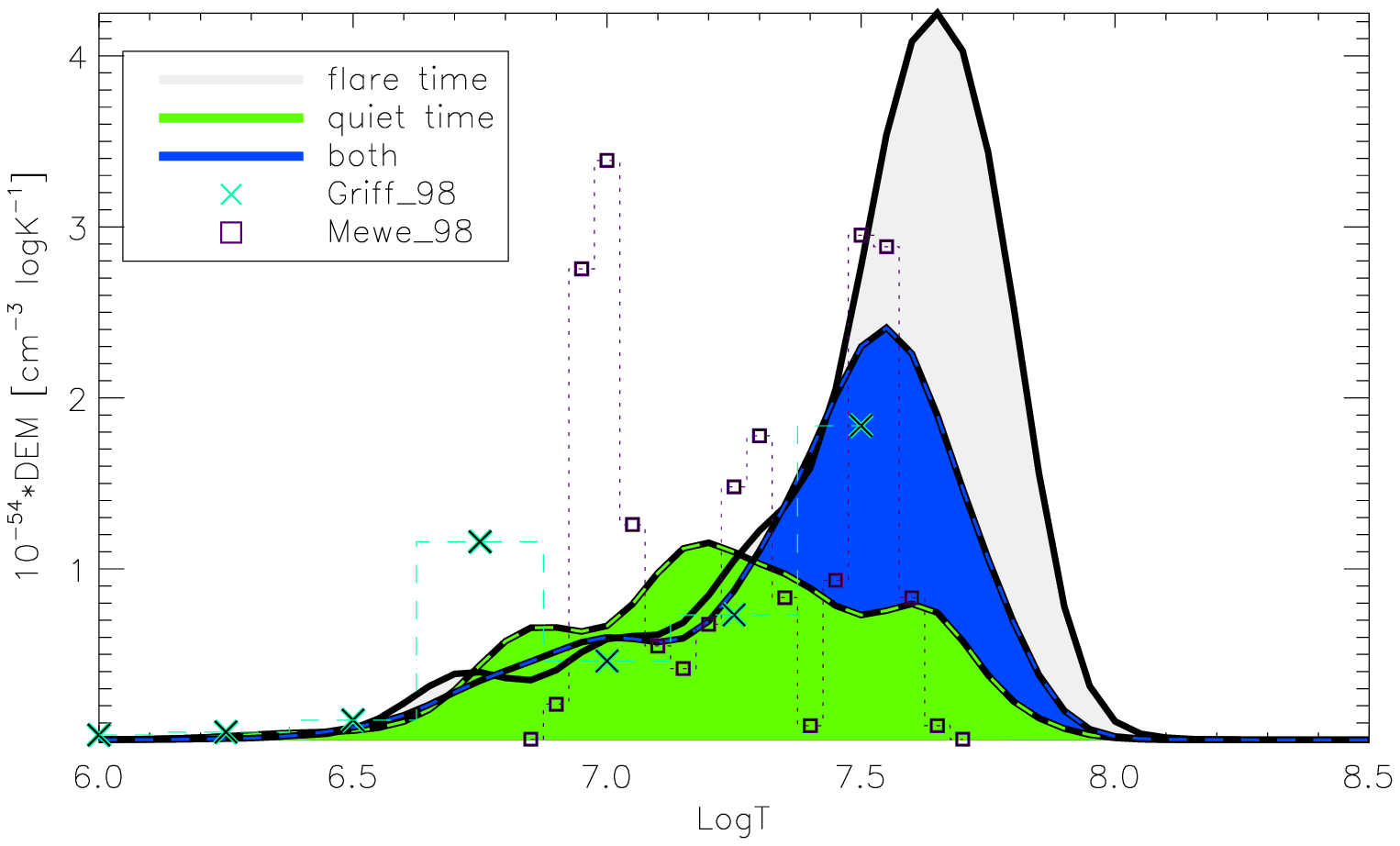}
 % \plotone{DEM.eps}
 % \plotone{DEM.pdf}
  \caption{\CapThree}
  \label{fig:dem}
  \end{figure}

Figure~\ref{fig:dem} shows our DEM fits as smooth curves filled with
gray shading; the quiescent state is medium gray, the flare state is
light gray, and the fit to the entire observation has the darkest
shading.  The wiggles in the thick outline curves and the deviations
between them between $\log T=6.5$ to 7.3 are characteristic of the
uncertainties in the data and models and are not significant. We see
two components in the temperature structure.  Before the flare, the
DEM rises gradually from a very low emissivity below $\log T=6.5$ to a
peak of about $10^{54}$ near $\log T=7.2$, then gradually declines,
becoming negligible above $\log T=7.9$. During the flare, a hot
component (from about $\log T=7.3$ to $8.0$) is present with a peak
value of about four times the maximum quiescent DEM. The lower
temperature region is largely unchanged by the flare.  This behavior
implies that the flare added a discrete source of emission, and that
the cooler source was constant.  The fit for the total exposure lies
between the two extremes, as it should for a time-weighted average of
flare and non-flare line fluxes.
                                \cbstart % refreply begin 2 [2001.05.09]
The addition of a hot component to the DEM with a relatively unchanged
cool component is a common characteristic for flares on RS~CVn
stars. \citet{Mewe1997} analyzed a flare on II~Peg seen with ASCA, and
found that the flare contribution could be characterized by the
addition of a single hot component. \citet{Gudel1999} concluded that
the cool DEM in UX~Ari was constant during a flare. \citet{Osten2000}
noted a slight enhancement in the cool plasma during a flare onset in
$\sigma^2$~CrB, which otherwise maintained a constant level of emission.
                                \cbend % refreply end 2 [2001.05.09]

We compare our emission measure with that derived by \citet{GRIFF98},
based on EUVE spectra.  We need to apply two scaling factors. First,
HIPPARCOS significantly revised the distance determination to II~Peg
from 29 pc to 42 pc \citep{HIPP1997}.  Second, \citet{GRIFF98} assumed
Solar abundances, but the EUVE spectrum is dominated by iron lines.
Since the iron abundance is in fact much below Solar (see below),
their emission measure must be scaled accordingly.  Applying these
factors, we find that their emission measure approximately follows our
mean curve, except for the large peak at $\log T=6.8$, as is shown in
Figure~\ref{fig:dem}.

\citet{Mewe1997} fit emission measure distributions to EUVE and ASCA
spectra, which we over-plot on Figure~\ref{fig:dem}.  Their structure
appears more peaked than ours.  While their flare DEM (the peak at
highest $\log T$) is comparable to ours, the lower temperature
structure is very different.  Integrated values are similar.

\begin{figure}
   \figurenum{4}
  \epsscale{1.0}
  \plotone{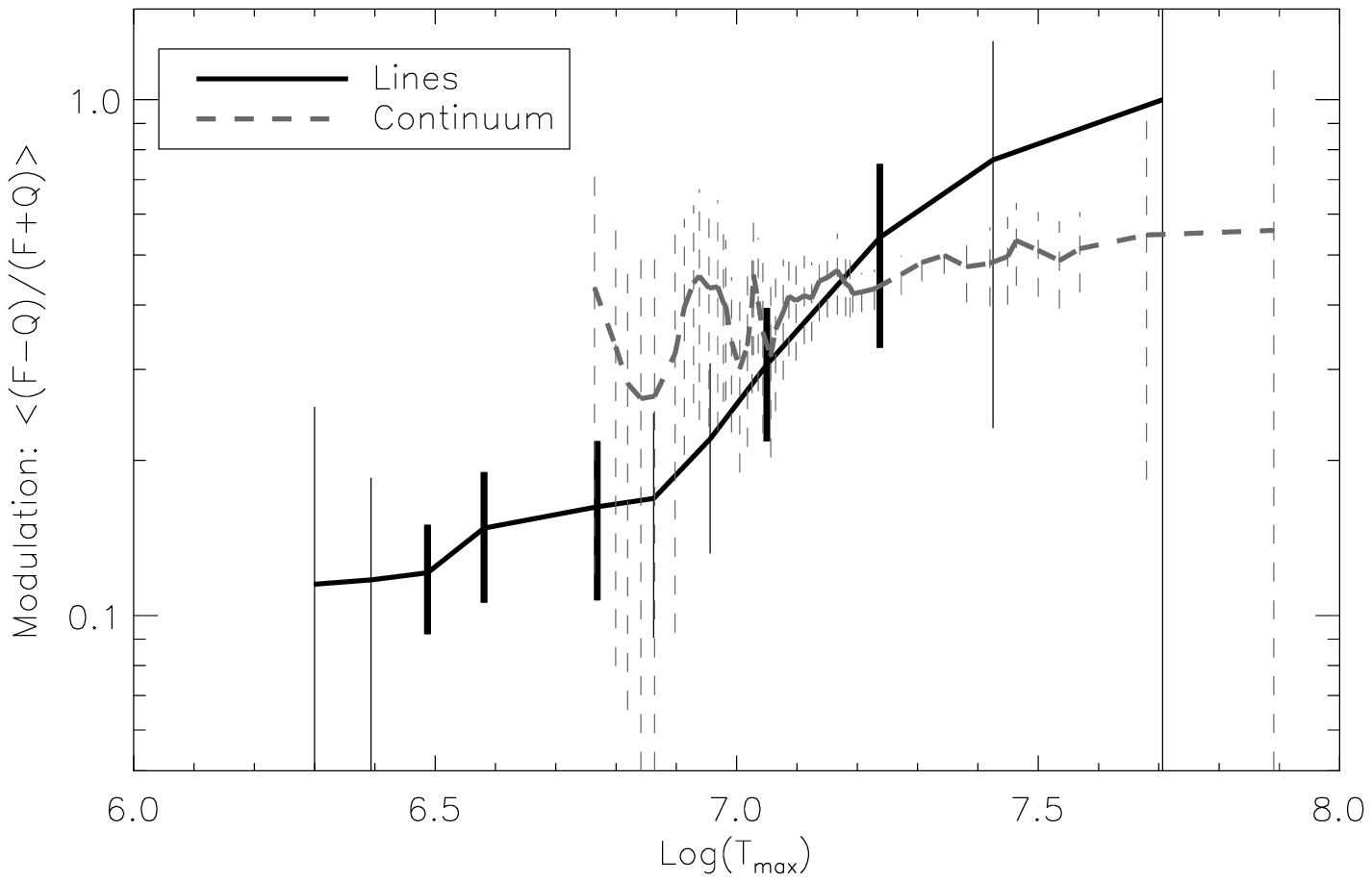}
 % \plotone{Line_Fluxes_b.eps}
 % \plotone{Line_Fluxes.pdf}
  \caption{\CapFour}
  \label{fig:linefluxes}
  \end{figure}

The high temperature fit region is primarily determined by high
ionization states of Fe, S, Si, and Ar.  The lower ionization states
of iron ({\sc xvii-xxi}) do not change much during the flare, but
lines from Fe~{\sc xxii-xxv} increased significantly. Other ions, such
as O~{\sc viii} and Ne~{\sc x}, whose emissivities peak at relatively
low temperatures ($\log T=6.5-6.8$) but also have extremely long tails
to very high temperature, also increased in strength.
Figure~\ref{fig:linefluxes} shows the flux modulation between
flare and quiescent states.  The modulation is defined as the ratio of
the flux difference to the flux sum for flare and quiescence.  For no
change, the modulation is 0.0, and for 100\% change, 1.0 (in the sense
of flare minus quiescent). We grouped lines by $\log T_\mathrm{max}$
to form mean fluxes before computing the modulation. The increase
in modulation
with $T_\mathrm{max}$ is clear.  In particular, note
that Fe~{\sc xxv} ($\log T_\mathrm{max}=7.7$) is not detected in the
pre-flare state, Fe~{\sc xxiv} ($\log T_\mathrm{max}=7.3$) changed
appreciably, but Fe~{\sc xvii} ($\log T_\mathrm{max}=6.7$) lines
change little.
                                \cbstart % refreply begin 4 [2001.05.09]
To explicitly quantify the changes for some lines, we tabulate
pre-flare and flare fluxes in Table~\ref{tab:fluxmod}.  The values
tabulated are representative of the modulations plotted, but are not
precisely the same line groups, so actual values will differ slightly
from those plotted.
                                \cbend % refreply end  4 [2001.05.09]

The continuum can also be used to constrain the temperature
distribution. In Equation~\ref{eq:dembasic}, $G(T)$ in this respect is
the continuum emissivity integrated over a given bandpass. Given the
form of the continuum emissivity and its slow variation with
temperature, the solution is not unique: the same flux could be
obtained from a flat or sharply peaked emission measure distribution.
Fitting the continuum in the same way as line fluxes is not practical.
Instead, continuum-band flux ratios can be used to estimate
temperatures, assuming that emission is dominated by a single
temperature.  We use the flux ratio of a 2--6 \AA\ band to the 7--8
\AA\ band from the APED model continuua, and compare the theoretical
ratios to observed ratios in flare and non-flare states.  The ratio
clearly changes, and is consistent with the DEM fits, indicating a
change from $\log T$ of 7.3 to 7.6 between quiescent and flare states.
This is not a unique result: using a longer wavelength band of 10--12
\AA, for example, systematically shifts derived temperatures to lower
values.  Continuum bands are thus qualitatively useful, but cannot
easily quantify the temperature distribution.

We computed a continuum modulation between flare and quiescent states
similarly to line fluxes. The continuum fluxes were taken from the
line plus continuum fits, and for a characteristic $\log T$, we
                                \cbstart % refreply begin 5 [2001.05.09]
computed a pseudo-temperature for each wavelength ($T =
hc/k\lambda$).  This is a crude parameterization of the temperature
dependence of the shortest wavelength continuum contribution, since
higher temperatures lead to stronger short-wavelength continuum.
                                \cbend % refreply end 5 [2001.05.09]
The continuum modulation is shown on
Figure~\ref{fig:linefluxes}. It increases very weakly to higher
temperature, with a mean value of about 0.5, which means that the
flare state had about three times the continuum flux as the quiescent
state.

The $\log$ integrated emission measures are $53.9$ and $54.3$ (for
volume in $\mathrm{cm^{-3}}$) for the quiescent and flare states,
respectively.  The corresponding observed fluxes and luminosities for
the 1--25 \AA\ wavelength range are 0.52 and 1.3 $\times
10^{-10}\mathrm{ergs\,cm^{-2}\,s^{-1}}$, and 1.1 and 2.9 $\times
10^{31}\mathrm{ergs\,s^{-1}}$, respectively.  These luminosities are
comparable to those obtained by \citet{Covino2000} but about twice
times that of \citet{Mewe1997} (after the latter were scaled to the
revised distance of 42 pc, a factor of two in luminosity).

\subsection{Elemental Abundances}

Relative elemental abundances are crucial to the fit of a DEM to line
fluxes. Absolute abundances relative to hydrogen can be deduced by
adjusting the line-to-continuum ratio.  The values we obtain are given
in Table~\ref{tab:abund}.  The uncertainties are determined by {\em ad
  hoc} perturbation of an abundance, generating a model spectrum, and
comparing it to the observed spectrum.  For the stronger lines (Ne,
O), a change of 10--15\% could be easily detected.  For elements with
only weak lines, such as Ar~{\sc xvii} or S~{\sc xv}, the
uncertainties are as large as a factor of two.

We also performed trial runs of the abundance fitting of line
fluxes as we did for the temperature distribution.  For several
trial DEM distributions, we assumed a set of trial abundances,
evaluated model line fluxes, introduced noise, and then fit to derive
the DEM and abundances. Variances in the derived trial abundances are
consistent with the above subjective inspection of models

Iron is extremely underabundant, at about 0.15 times Solar (-0.8 dex).
Neon is overabundant by about a factor of two.  
                                \cbstart % refreply begin 6a [2001.05.09]
\citet{Ottman1998} modeled the photospheric lines of iron, magnesium,
and silicon in \iip.  Their abundances relative to Solar were about
0.6, 0.7, and 0.7, respectively (-0.22, -0.15, and -0.15 dex).
\citet{Berdy98} determined a comparable photospheric metallicity of
0.3--0.5 times Solar.  Our coronal abundance for iron is about four
times lower, while Mg and Si are within 70\% of their values and are
within our respective uncertainties.  \citet{Mewe1997} fit EUVE and
ASCA spectra of \iip, and they required variable, non-Solar
abundances.  Our iron abundances are comparable, but values for other
species differ by about a factor of 3--4, though the neon to oxygen
ratio is similar.  The differences are probably due to uncertainties
in spectral fitting to the previous low-resolution data.
                                \cbend % refreply end 6a [2001.05.09]

The DEM fit gives neon
to iron ratios relative to the Solar ratio of $22\pm6$ (pre-flare) to
$17\pm5$ (flare) or $16\pm4$ (total spectrum).  Fe~{\sc xvii} and
Ne~{\sc ix} have nearly identical theoretical emissivity profiles:
they are both sharply peaked, and their temperatures of maximum
emissivity are separated by less than 0.1 in $\log T$.  Hence, the
flux ratios of Fe~{\sc xvii} to Ne~{\sc ix} should only depend upon
their relative abundances and atomic parameters.  We compared flux
                                \cbstart % refreply begin 6b [2001.05.09]
ratios of Ne~{\sc ix} 13.446 \AA\ resonance line, since it is
relatively unblended, to the Fe~{\sc xvii} line at 15.01.  The flux
ratios and APED emissivities imply that the neon to iron abundance
ratio (for the total spectrum) is $14\pm2$ times the Solar ratio. For
the pre-flare and flare intervals, respectively, we obtain the ratios
$14\pm4$ and $16\pm5$.  The line-ratio and DEM values are all
identical, given the statistical uncertainties.
                                \cbend % refreply end 6b [2001.05.09]

\subsection{Densities}

The helium-like triplets of O~{\sc vii}, Ne~{\sc ix}, Mg~{\sc xi}, and
Si~{\sc xiii} which are detected and resolved into the resonance ({\em
  r}), intercombination ({\em i}), and forbidden ({\em f}) lines by
the HETGS.  These lines are useful because the flux ratio of $f/i$ is
primarily density-sensitive, and $(f+i)/r$ is mainly temperature
sensitive \citep{Gabriel69, Pradhan82, Porquet2000}. The ratios are
not dependent upon the ionization balance since they are from the same
ion, and since they are close together in wavelength they are
relatively insensitive to calibration uncertainties.  The critical
density, above which the ratio, $f/i$, drops from a nearly constant,
low-density limit, rises by about a decade for each of O~{\sc vii},
Ne~{\sc ix}, Mg~{\sc xi}, to Si~{\sc xiii}, starting near $10^{10}
\mathrm{cm^{-3}}$ for oxygen to $10^{13} \mathrm{cm^{-3}}$ for Si,
spanning an interesting range of expected coronal densities.  The
He-like triplets of S~{\sc xv}, Ca~{\sc xix}, Ar~{\sc xvii}, and
Fe~{\sc xv} lines are also in the HETGS band, but are either weak or
unresolved.  Their critical densities are also well above the coronal
range.

We have determined the limits in density implied by the APED models
for our measured line fluxes.  Since the lines did not change
substantially during the flare, we used the flux integrated over the
entire observation to achieve the best signal.  We also only consider
temperatures near the peak emissivity times emission measure, which
limits the range to roughly $\log T$ 6.0--7.0 for O~{\sc vii}, Ne~{\sc
  ix}, and Mg~{\sc xi}.  The 68\% confidence limits in $\log N_e$ for
oxygen are 10.6--11.6 $[\log \mathrm{cm^{-3}}]$.  Ne~{\sc ix} gives
11.0--12.0, with a barely constrained lower limit. The range for
Mg~{\sc xi} is 12.8--13.8.  The respective 90\% limits for O~{\sc
  vii}, Ne~{\sc ix}, and Mg~{\sc xi} are 10.3--12.4, $<12.3$, and
$>11.8$. Si~{\sc xiii} density is unconstrained.

The neon and oxygen density limits overlap, while magnesium and neon
are marginally inconsistent.  There are plausible systematic affects,
such as line blends and continuum placement. The Ne~{\sc ix}
intersystem line (13.55 \AA) is blended with Fe~{\sc xix} (13.52 \AA),
but two peaks are resolved and are easily fit with two components.
Unaccounted blends in the Mg~{\sc xi} $i$ could lower the measured
$f/i$ ratio resulting in an erroneously higher density.  The observed
Mg~{\sc xi} $i$ flux is significantly greater than the predicted model
(see Figure~\ref{fig:specdetailb}, 9.23\AA region).  

                                \cbstart % refreply begin #ours [2001.05.09]
We believe the discrepancy is caused by blending of the Ne~{\sc x}
Lyman series with the Mg~{\sc xi} lines. Levels above $n=5$
are not included in the APED models, and they are probably responsible
for the clear excess of data over prediction in the 9.2-9.5 \AA\ 
range.  In particular, the Ne~{\sc x} Lyman series lines from upper
levels 6--10 fall at 9.362, 9.291, 9.246, 9.215, and 9.194 \AA.  These
will have to be included in models if we are to obtain reliable
Mg~{\sc xi} triplet ratios, especially in stars with enhanced neon
abundance.
                                \cbend % refreply end #ours [2001.05.09]

An increase in the Mg~{\sc xi} $f/i$ ratio by about 50\% would put the
density lower-limit at about 12.3, while doubling the ratio would make
the lower-limit unconstrained. It is difficult to reconcile densities
between Mg and O without a factor of two error in the Mg ratio; given
the model uncertainties and the difficulty of fitting the blends, such
an error is likely.

For two-sigma confidence intervals, O, Mg, and Ne together imply
consistent logarithmic density near 11.8--12.3.  However, there is no
{\em a priori} reason to expect density to be constant with
temperature, and hence the same for each ion.

\subsection{Resonant Scattering}\label{sec:scatt}

The ratio of the Fe~{\sc xvii} $\lambda15.265$ ($2p^6\,^1S_0$ ---
$2p^53d\,^3D_1$) to its resonance line neighbor at $\lambda15.014$
(upper level $2p^53d\,^1P_1$) is a measure of density and geometry,
since the latter has a shorter mean free path.  \citet{Saba99} have
summarized Solar measurements and theoretical values for the ratio and
concluded that opacity is a factor in the Sun; theoretical
calculations and Solar measurements give ratios of $0.25\pm0.04$ and
$0.49\pm0.05$, respectively.  Recent laboratory measurements obtained
values in the range $0.33\pm0.01$ \citep{Brown98}.  \citet{Laming2000}
measured these as well as the nearby 17 \AA\ lines sharing the same
lower level, and compared to theory.  Their observed ratio was
$0.40\pm0.02$, and the ratio of the 17 \AA\ pair to the resonance line
at 15 \AA\ was $0.9\pm0.1$, for a beam energy of 1.25 keV.  Both
Brown and Laming et al.\ suggested that opacity effects in the Sun
have been overestimated, since the improved measurements have raised
the ratios towards the Solar values, and because there are physical
processes other than collisional excitation and radiative decay which
have not yet been accounted for in the theoretical calculations.

We have measured a 15.26 to 15.01 flux ratio of $0.48\pm0.14$ in the
HETGS spectra, which is as large as the Solar values, and larger than
the value of $0.26\pm0.1$ determined for Capella
\citep{Brinkman2000, Canizares2000}. The uncertainty, however, is large
enough that the result is not inconsistent with the other values. 
We see no direct evidence of opacity in Fe~{\sc xvii} in II~Peg. 

Our ratios of the 17 \AA\ lines to the 15 \AA\ line is $2.1\pm0.4$,
which is twice the laboratory value of $0.9\pm0.1$.  This ratio is
weakly temperature dependent; the APED ratio ranges from 0.8 at $\log
T=7.2$ to 1.3 at 6.1.  There is some systematic incurred in our
measured value from continuum placement, but not a factor of two.  If
we had used these lines for our relative neon-to-iron abundance
determination, we would have obtained a value a factor of two lower.
(The lines were used implicitly in the DEM fitting, but their affect
is diluted by the presence of other lines of Fe.)  \citet{Laming2000}
discuss some possibilities for ``small'' contributions to the upper
levels which produce the 17\AA\ lines, such as recombination from
Fe~{\sc xviii} into excited levels, but none can explain the large
discrepancy between various Solar and stellar observations, which they
show in their Figure~3.  We have neither an atomic nor plasma physics
explanation for the anomalous relative strength of Fe~{\sc xvii}
17\AA\ lines.

We have also looked at the ratios of the Ly-$\alpha$-like series for
O~{\sc viii}, Ne~{\sc x}, and Si~{\sc xiv}.  None differ significantly
from the APED theoretical ratios.  The largest difference is for
O~{\sc viii}; the $\beta:\alpha$ ratio of $0.19\pm0.015$ (68\%
confidence) and the theoretical value is 0.16.  Hence, we have no
direct evidence of scattering in O~{\sc viii} Ly-$\alpha$.  The ratio
is not affected by interstellar absorption, since the column density
require to change the $\beta:\alpha$ to exactly match the theoretical
value would make II~Peg invisible to EUVE.  
                                \cbstart % refreply begin 7 [2001.05.09]
Estimates have been made from EUVE spectra of a few times
$10^{18}\,\mathrm{cm^{-2}}$ \citep{GRIFF98}, which we will adopt.
This amount has negligible affect on the X-ray spectrum.
                                \cbend % refreply end 7 [2001.05.09]
None of the Ly-series ratios for oxygen or neon
changed significantly during the flare.

                                \cbstart % refreply begin 3 [2001.05.09]
\subsection{Line Profiles and Shifts}\label{sec:lineprofiles}

Line shifts and shapes can be important diagnostics of plasma
dynamics.  Flares may have upflows or downflows of up to 400
$\mathrm{km\,s^{-1}}$, which would be resolvable by HETGS.  However,
we observe the flare emission against the background quiescent
emission, and the flare emission was predominantly in the continuum,
though lines did increase somewhat in flux. We did not see any
significant line shifts during the flare, which we inspected by
differencing the pre-flare spectrum from the flare spectrum. Small
shifts would show as differential profiles, but we only saw the small
change in line flux.

Line shapes are more difficult to assess, especially if broadening is
comparable to the instrumental resolution.  A very good calibration of
the instrumental profile is needed.  Profile fits suggest that
O~{\sc viii} 19\AA\ is slightly broader than instrumental, but we are
not yet confident enough on the calibration of the intrinsic profile
at this level.  The calibration is being improved, and we will apply
it when available.
                                \cbend % refreply end 3 [2001.05.09]

\section{Discussion}
\label{sec:disc}

\subsection{Abundances and Temperature Structure}
\label{sec:abundtemp}

There has been much discussion and controversy regarding low metal
abundances in stellar coronae.  Previous low-resolution investigations
found reduced abundances of some elements, such as for
                                \cbstart % refreply begin 9 [2001.05.09]
UX~Ari \citep{Gudel1999},
$\sigma^2$~CrB \citep{Osten2000}, and II~Peg \citep{Mewe1997}.
                                \cbend % refreply end 9 [2001.05.09]
\citep[Also see reviews by][]{White1996,  Pallavicini1999}.  
Without being able to resolve individual lines,
counter-arguments have criticized over-simplistic emission measure
models which fit only two temperature components, limitations of
emissivity models in which a pseudo-continuum from many weak lines
would artificially reduce line-to-continuum ratios, enhanced helium
abundance leading to stronger continuum \citep{Drake1998}, ignorance
of photospheric abundances, or calibration uncertainties.
\citet{Drake96} applied the term, ``MAD'' (for Metal Abundance
Deficient) for the low-abundance objects.  Much of the interest in
abundances has been driven by empirical evidence of the ``First
Ionization Potential'' (FIP) affect in the Sun 
\citep{Feldman90,  Laming1996} in which easily ionized elements are
over-abundant in 
the corona. This is generally the opposite of what is inferred in
other stars: the FIP affect enhances iron in the Solar corona relative
to other elements.
                                \cbstart % refreply begin 8 [2001.05.09]
\citet{FeldmanLaming2000} give a thorough review of the state of
abundance determinations in the coronae of the Sun and other stars.
                                \cbend % refreply end 8 [2001.05.09]

With Chandra spectra, we are now able to settle some of these
questions. \citet{Drake2001} have found neon to be much enhanced
relative to iron in HR~1099.  II~Peg is similar: it is clearly
deficient in iron, and neon is enhanced.  This is a not a simple reflection
of the photospheric abundances.  
                                %\cbstart % refreply begin 6 [2001.05.09]
% stuff moved to 4.2
                                %\cbend % refreply end 6 [2001.05.09]

While we have improved the quality of abundance determinations in
II~Peg, we cannot explain them.  II~Peg is somewhat MAD, but has high
neon.  It definitely does not show the FIP affect, but much the
opposite (we tabulate the FIP in Table~\ref{tab:abund}). The iron
deficiency is quite strong, and there are large discrepancies between
some observed ratios and the theoretical or laboratory values.
\citet{Brinkman2001} reported an inverse FIP affect in HR~1099 as
determined from XMM-Newton Reflection Grating Spectrometer data.  From
HETGS spectra of HR~1099, \citet{Drake2001} determined abundances
similar to Brinkman's values. Neither HR~1099 nor II~Peg show a truly uniform
FIP affect.  At the lowest ionization potentials, there is a factor of
several spread between the low Fe and moderate Mg and Si abundances.

                                \cbstart % refreply begin 9 [2001.05.09]
Our preliminary work on other stars with HETGS spectra hints that
there is a range of iron abundance from very low in II~Peg, to
moderate depletion AR~Lac and TY~Pyx, which seem to have higher Fe/Ne
abundance ratio than HR~1099.  Capella appears to be at the
``normal'' end of the distribution, with neither strong iron nor neon
deviations from cosmic abundances \citep{Audard2001}.
                                \cbend % refreply end 9 [2001.05.09]

The temperature structure we obtain is much different from previous
DEM determinations.  Ours is much smoother.  While some smoothing has
been imposed to make the fit better behaved, we should have been able
to resolve features as sharp as found by \citet{GRIFF98} and
\citet{Mewe1997}. Our simulations showed that we could fit peaks with
$FWHM\sim0.15$ dex in $\log T$.  We suspect that the improved
resolution and spectral coverage spanning a greater range in
ionization states and elements are key factors. 
                                \cbstart % refreply begin 12 [2001.05.09]
Further studies are required to determine the affects of fitting
methods, spectral resolution, and range of model space provided by
spectral features on resulting DEM distributions.
                                \cbend % refreply end 12 [2001.05.09]

\subsection{Flare Models}

It is possible to model flare light-curves to derive constraints on
the loop sizes, magnetic fields, and densities.  \citet{Kopp1984} have
formulated a model for two-ribbon flares based on detailed Solar
observations, which \citet{Poletto1988} extended to other stars.
Their model describes the conversion of magnetic energy to X-rays via
reconnection of rising loops, assuming the flare occurs at the site of
reconnection at the top of the rising loops.  This model has been
frequently applied to ultraviolet and X-ray flares on RS~CVn stars
\citep{Gudel1999, Osten2000}.  A complementary approach can be found
in \citet{vdOord1988} and subsequent papers
\citep[e.g.][]{vdOord1989,vdOord97}, including application to \iip\ 
\citep{Mewe1997}, in which conductive and radiative energy balance are
applied to the flux and temperature profiles.

There are a large number of parameters in these models, some poorly
constrained, such as the fraction of energy emitted in X-rays, the
magnetic field, the number of loops, the shape of loops, and the
electron density.  The observable quantities are the shape of the
light-curve, the light-curve amplitude, and the spectrum of the
plasma.  The models are most constrained by the decay phase under the
assumption that heating has stopped.  Even this simplification is
suspect, however, since flare light-curves have shown extremely long
or structured decay \citep{Osten2000}.

We do not observe enough of the flare decay in this observation to
determine two of the fundamental parameters in the Kopp \& Poletto
model: the loop size, and the time constant.  The decay function
\citep[see equations 2--6 in][]{Gudel1999} differs most for different
loop sizes at about 50 ks into the flare, whereas we have observed
only about 25 ks.  We did not observe enough of the decay to determine
a cooling profile from the spectrum.  There is no density diagnostic
available in the net flare spectrum above the constant background
coronal emission.  Better diagnostics will come when a very large
flare is observed over the right time interval, which is bright enough
to measure density diagnostics in spectral lines.

We have compared the two-ribbon reconnection model energy profile to
the light-curve we have.  The smooth line plotted in
Figure~\ref{fig:lcbasic} shows the arbitrarily normalized model for a
time-constant of 65 ks and a point-like flare.  The top portion of the figure shows the counts integrated
over all diffracted photons, while the bottom shows a narrow band from
6.8-8.2 \AA.  The normalization factor for the latter requires about a
20\% relative difference from the broad-band curve, indicating that
there is perhaps some chance of constraining the model with continuum
band information.  We will pursue low-resolution light curve
quantitative flare diagnostics in future work.

                                \cbstart % refreply begin 10 [2001.05.09]
The time-constant is not a characteristic cooling time, nor the
exponential decay constant, but is the Kopp \& Poletto reconnection
time scale parameter related to loop dynamics and continuous heating.
If we had more of the decay profile, we might be able to assess the
relative affects of continuous heating (reconnection), or whether the
probably long decay is due to less efficient conductive and radiative
cooling due to larger, lower density structures than seen on the Sun.
\citet{Golub1989} list some characteristic cooling times for different
density structures (their Table~V), which span several orders of
magnitude from 500 to $10^5$ seconds. Hence, it is possible to have
long-lived flares without continuous heating if densities are low.
                                \cbend % refreply end 10 [2001.05.09]

We do not see any evidence of impulsive behavior before the obvious
flare onset at about 24 ks, but this could be easily masked by the
background quiescent emission. Impulsive events are also likely
to be harder than the response range of the HETGS, since these
precursor flares are characterized by rapid variability at energies
greater than 20 keV.

There is a hint that Ne and Fe are enhanced during the flare (see
Table~\ref{tab:abund}), but it is $\le 2\sigma$ affect, and not
conclusive that flare abundance variations occur in II~Peg, as they do
in the Sun.

\subsection{Loop Sizes}

Given our emission measure determination and density estimates, we
estimate loop sizes under the simplifying assumption of uniform cross
section, semicircular loops intersecting a plane (a hemi-toroidal
loops).
                                \cbstart % refreply begin 11 [2001.05.09]
We define $L$ as the loop length from foot point to foot point along
the loop axis, and $h$ as the height defined from the midpoint of the
segment connecting the centers of the circular bases to the center of
the loop cross-section (i.e., the radius of the axis of the loop).
Thus, $L=\pi h$.  We define $\alpha$ as the ratio of the loop
cross-sectional radius to the loop length.  The loop volume is then
$V = \pi^4 \alpha^2 h^3$.  Noting that the volume emission measure, $VEM$,
is approximately $N_e^2\,V$, we can write the loop height as a
function of $VEM$ and $N_e$
                                \cbend % refreply end 11 [2001.05.09]
in units of the stellar radius, $R_*$, for an ensemble of identical
emitting loops as
\begin{equation}
  \label{eq:loopheightgenl}
  h =
  \pi^{-4/3}\,N_{100}^{-1/3}
  \alpha_{0.1}^{-2/3}
  (VEM)^{1/3} n_{e}^{-2/3} R_{*}^{-1},
\end{equation}
where $N_{100}$ is the number of loops divided by 100 (an arbitrary
normalization), and $\alpha_{0.1}$ is the ratio of loop radius to loop
length in units of 0.1 (a number typically used for Solar loops).

For our determinations of $VEM=7.9\times10^{53}\,\mathrm{cm^{3}}$ (for
quiescent state) and $n_e\sim10^{11} \,\mathrm{cm^{3}}$, a stellar
radius of $3R_\sun$ \citep{Berdy98}, $N_{100}=1$, and $\alpha=1$, we
find that $h = 0.05$.  The loops are small compared to the stellar
radius.  If we compute the height for the flare excess $VEM$ and
assume a single loop, then it could be 0.25 of the stellar radius for
the same density.  If flare density were enhanced by a factor of 10,
then the loop size remains at 0.05.  It may be possible to detect
loops of this size in other systems via X-ray light curves of
eclipses.  We are pursuing such observations.

\subsection{Model Spectrum}

A final test of any model is a prediction of the observed data,
especially observed features which were not used in determination of
the model.  We use the derived emission measure and abundances for the
entire duration of the observation to model a spectrum (using the APED
in ISIS) and fold this through the instrumental effective area and
line-spread-function.
Figures~\ref{fig:specdetail}-\ref{fig:specdetaild} show the counts
spectrum and folded model.  There are features which are clearly not
in the APED database, or have poorly determined wavelengths or
emissivities, or where there are larger calibration errors.  This
comparison does not show that the model is necessarily ``correct'',
but that it is not inconsistent with the observation.  It is actually
a very good match, but there are clearly regions where better fits
should be pursued, and discrepancies between models and data resolved.

\section{Conclusions}

II~Pegasi is a very interesting and much studied RS~CVn star,
uncomplicated by a secondary star's spectrum.  We have confirmed that its
corona is indeed iron deficient, and deficient relative to the
photosphere. We have also shown that neon has a substantially enhanced
coronal abundance.  Our temperature structure is much smoother than
prior studies which does not seem to be an artifact of DEM modeling.  A
moderate sized flare enhanced the hot component, as evidenced
primarily in the continuum.  The flare is consistent with a Solar
two-ribbon model, but not enough of the decay was observed to provide
physical constraints.  Loop sizes are fairly small, about 5\% of the
stellar radius, but this is larger than estimates for Capella
\citep{Canizares2000} and perhaps enough to encourage X-ray light
curve studies of eclipsing systems.

\acknowledgements Work at MIT was supported by NASA through the HETG
contract NAS8-38249 and through Smithsonian Astrophysical Observatory
(SAO) contract SVI--61010 for the Chandra X-ray Center (CXC).  We
thank all our colleagues who helped develop and calibrate the HETGS
and all members of the Chandra team, in particular the CXC programming
staff. We especially thank John Houck for his support of ISIS, Dan
Dewey for a critical review of the manuscript, and Nancy Brickhouse
for providing insight regarding the Mg~{\sc xi} ratios.

%\input{Bibliography}

%%%%%%%%%%%%%%%%%%%%%%%%%%%%%%%%%%%%%%%%%%%%%%%%%%%%%%%%%%%%%%%
%%%%%%%%%%%%%%%%%%%%%%%%%%%%%%%%%%%%%%%%%%%%%%%%%%%%%%%%%%%%%%%
%%%%%%%%%%%%%%%%%%%%%%%%%%%%%%%%%%%%%%%%%%%%%%%%%%%%%%%%%%%%%%%
%%%%%%%%%%%%%%%%%%%%%%%%%%%%%%%%%%%%%%%%%%%%%%%%%%%%%%%%%%%%%%%

% the following is for use in the -ms (manuscript) form:

\clearpage 
\figcaption[Full_spec.eps]{\CapOne}
\figcaption[Lc_basic.eps]{\CapTwo}
\figcaption[DEM.eps]{\CapThree}
\figcaption[Line_Fluxes_b.eps]{\CapFour}
\figcaption[Spec_Detail.eps]{\CapFive}
\figurenum{5b}\figcaption[]{\CapFiveB}
\figurenum{5c}\figcaption[]{\CapFiveC}
\figurenum{5d}\figcaption[]{\CapFiveD}

%%%% commented out for manuscript submission

\onecolumn
\clearpage

\clearpage

\clearpage

\clearpage

\clearpage
  \begin{figure}
   \figurenum{5a}
  \epsscale{0.9}
%  \plotone{f5a.eps}
 % \plotone{Spec_detail_alt_A.ps}
 % \includegraphics[angle=-90]{Spec_detail_alt_A.ps}
$$\includegraphics[scale=0.95,angle=-90]{f5a_online.eps}$$
 % \caption{\CapFive}
  \caption{}
  \label{fig:specdetail}
  \end{figure}

\clearpage
  \begin{figure}
   \figurenum{5b}
  \epsscale{0.9}
%  \plotone{f5b.eps}
 % \plotone{Spec_detail_alt_B.ps}
 % \includegraphics[angle=-90]{Spec_detail_alt_B.ps}
$$\includegraphics[scale=0.95,angle=-90]{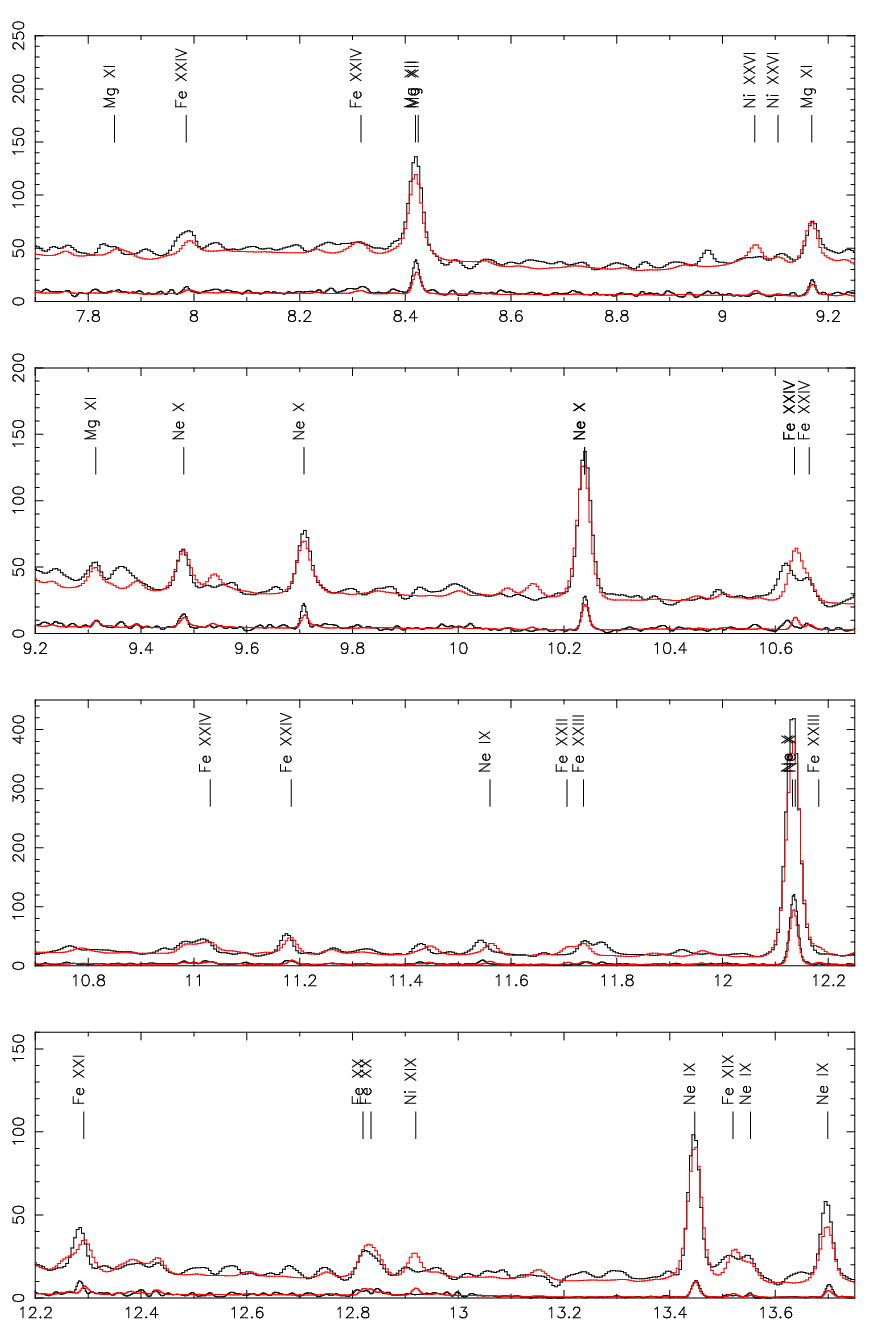}$$
 % \caption{\CapFiveB}
  \caption{}
  \label{fig:specdetailb}
  \end{figure}

\clearpage
  \begin{figure}
   \figurenum{5c}
  \epsscale{0.9}
%  \plotone{f5c.eps}
 % \plotone{Spec_detail_alt_C.ps}
 % \includegraphics[angle=-90]{Spec_detail_alt_C.ps}
$$\includegraphics[scale=0.95,angle=-90]{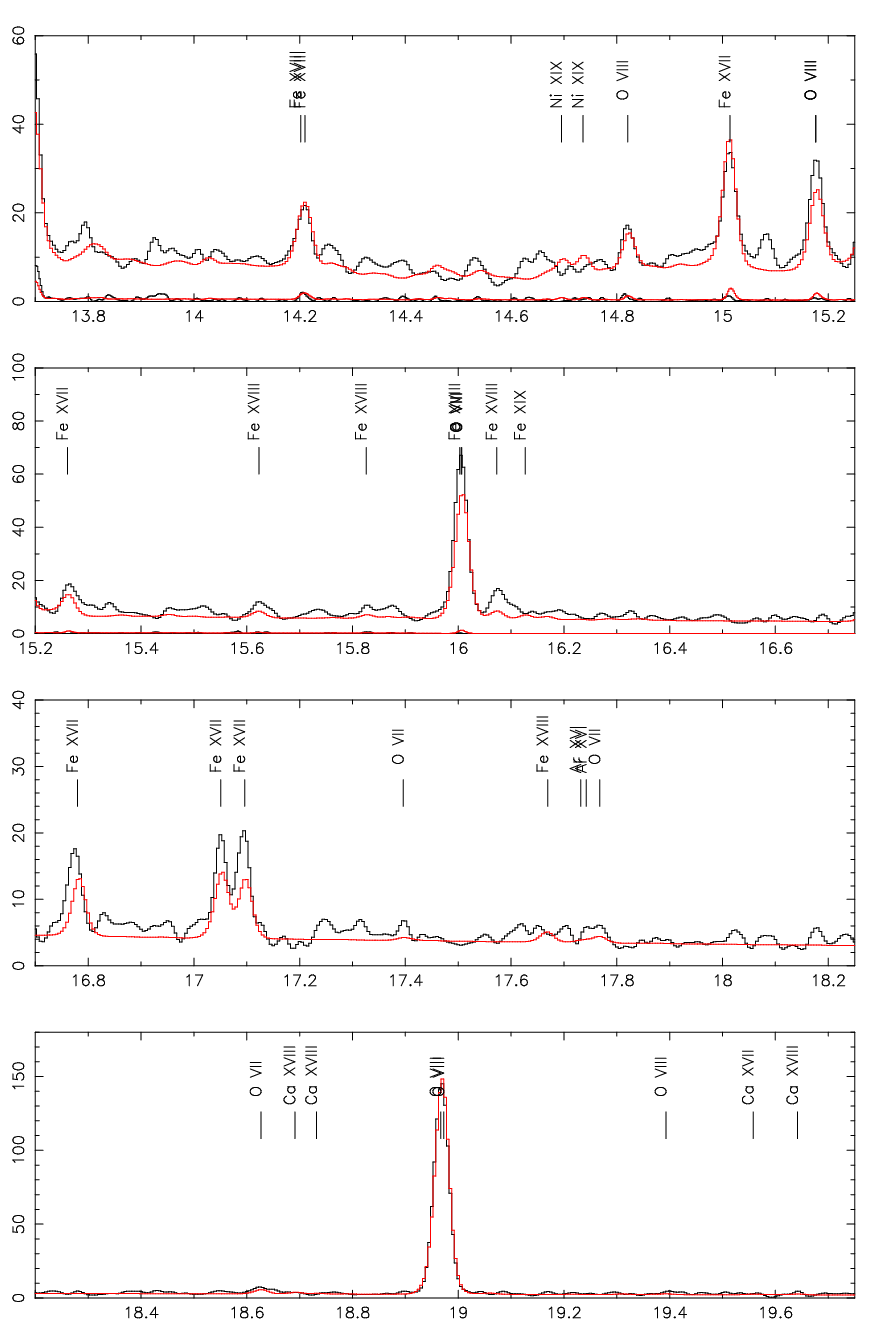}$$
 % \caption{\CapFiveC}
   \caption{}
  \label{fig:specdetailc}
  \end{figure}

\clearpage
  \begin{figure}
   \figurenum{5d}
  \epsscale{0.9}
%  \plotone{f5d.eps}
 % \plotone{Spec_detail_alt_D.ps}
 % \includegraphics[angle=-90]{Spec_detail_alt_D.ps}
$$\includegraphics[scale=0.95,angle=-90]{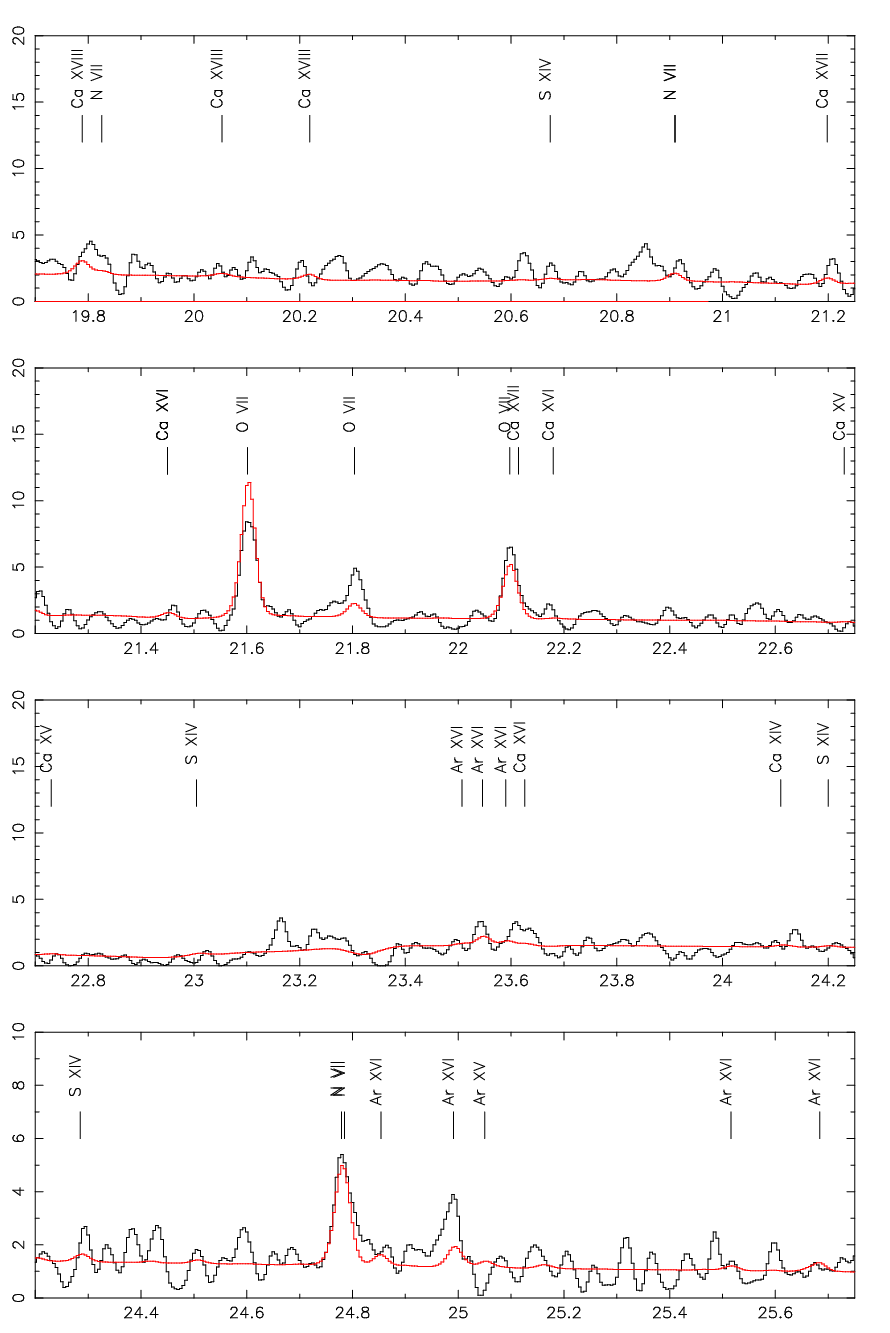}$$
 % \caption{\CapFiveD}
  \caption{}
  \label{fig:specdetaild}
  \end{figure}

%%%%%%%%%%%%%%%% 
 \clearpage
 {\small
 \begin{deluxetable}{rrrrrrrrc}
 \tablewidth{0pt}
 \tablecaption{Emission line measurements for entire spectrum.\label{tab:linelist}}
 \tablehead{
 \colhead{Line}&
 \colhead{$\lambda_t$\tablenotemark{a}}&
 \colhead{$\lambda_o$\tablenotemark{b}}&
 \colhead{$\sigma$}&
 \colhead{$f_l$\tablenotemark{c}}&
 \colhead{$\sigma$}&
 \colhead{$f_c$\tablenotemark{d}}&
 \colhead{$\sigma$}&
 \colhead{$\log T_{max}$\tablenotemark{e}}
 }
 \startdata
 Fe \sc xxv     & 1.8504&   1.857& 9.250e-03&    13.27&    12.84&   384.3&   156.8&  7.8\\
 Ar \sc xviii   & 3.7311&   3.734& 4.850e-03&    14.31&     5.49&  1192.4&    80.2&  7.6\\
 Ar \sc xvii    & 3.9491&   3.949& 9.500e-03&     6.75&     6.01&  1295.2&    89.2&  7.1\\
 Ar \sc xvii    & 3.9694&   3.969& 1.500e-02&    13.30&    26.01&  1295.2&    89.2&  7.1\\
 S  \sc xvi     & 4.7274&   4.727& 2.450e-03&    49.33&     8.92&  1280.8&    93.5&  7.4\\
 S  \sc xv      & 5.0387&   5.037& 1.450e-03&    30.21&     7.90&  1293.4&    76.1&  7.2\\
 S  \sc xv      & 5.1015&   5.095& 3.950e-03&    17.35&     7.75&  1293.4&    76.1&  7.2\\
 Si \sc xiv     & 5.2180&   5.221& 4.950e-03&    17.36&     8.22&  1381.5&   102.8&  7.2\\
 Si \sc xiv     & 6.1805&   6.180& 9.990e-05&    80.10&     5.33&  1483.5&    45.1&  7.2\\
 Si \sc xiii    & 6.6479&   6.648& 1.200e-03&    53.08&     4.65&  1476.9&    40.6&  7.0\\
 Si \sc xiii    & 6.6882&   6.685& 3.900e-03&     8.40&     3.74&  1476.9&    40.6&  7.0\\
 Si \sc xiii    & 6.7403&   6.740& 9.990e-05&    29.78&     3.86&  1476.9&    40.6&  7.0\\
 Fe \sc xxiv    & 7.9857&   7.987& 2.450e-03&    12.30&     3.23&  1435.0&    42.2&  7.3\\
 Mg \sc xii     & 8.4193&   8.420& 5.007e-05&    79.75&     4.92&  1412.6&    44.8&  7.0\\
 Mg \sc xi      & 9.1687&   9.170& 1.497e-04&    40.63&     4.85&  1557.8&    43.9&  6.8\\
 Mg {\sc xi}\tablenotemark{f}
                & 9.2312&   9.238& 2.750e-03&    10.17&     4.04&  1557.8&    43.9&  6.8\\
 Mg {\sc xi}\tablenotemark{f}
                & 9.3143&   9.315& 2.200e-03&    13.95&     4.31&  1557.8&    43.9&  6.8\\
 Ne \sc x       & 9.4807&   9.480& 1.497e-04&    38.20&     5.38&  1317.7&    72.6&  6.8\\
 Ne \sc x       & 9.7081&   9.706& 1.200e-03&    62.06&     5.20&  1321.8&    45.9&  6.8\\
 Ne \sc x       & 10.239&  10.240& 5.007e-05&   178.60&     8.46&  1188.1&    95.2&  6.8\\
 Fe \sc xxiv    & 10.636&  10.620& 5.007e-05&    47.21&     6.20&  1391.8&    55.1&  7.3\\
 Fe \sc xxiv    & 10.664&  10.660& 2.999e-04&    30.19&     5.88&  1391.8&    55.1&  7.3\\
 Fe \sc xxiv    & 11.184&  11.176& 1.400e-03&    68.96&     7.89&  1214.9&    88.1&  7.3\\
 Ne \sc ix      & 11.560&  11.545& 1.001e-04&    48.70&     7.98&  1408.1&    77.6&  6.6\\
 Fe \sc xxii-iii& 11.737&  11.743& 2.100e-03&    50.74&     7.93&  1423.4&    75.7&  7.1\\
 Ne \sc x       & 12.132&  12.132& 2.999e-04&  1110.30&    21.60&  1415.8&   104.5&  6.8\\
 Fe \sc xxi     & 12.292&  12.285& 2.999e-04&    74.76&    10.23&  1537.0&   103.3&  7.1\\
 Fe \sc xx      & 12.820&  12.821& 2.450e-03&    38.30&     9.04&  1375.1&   110.0&  7.0\\
 Fe \sc xx      & 12.835&  12.843& 2.450e-03&    35.97&     8.51&  1375.1&   110.0&  7.0\\
 Ne \sc ix      & 13.447&  13.446& 1.250e-03&   350.68&    19.84&  1283.8&   107.5&  6.6\\
 Fe \sc xix     & 13.516&  13.511& 2.550e-03&    72.40&     6.49&  1283.8&   107.5&  6.9\\
 Ne \sc ix      & 13.553&  13.550& 1.502e-04&    75.58&    12.10&  1283.8&   107.5&  6.6\\
 Ne \sc ix      & 13.699&  13.700& 9.966e-05&   203.29&    17.78&  1498.7&   123.5&  6.6\\
 Fe \sc xviii   & 14.210&  14.210& 1.550e-03&    69.39&    15.21&  1471.1&   134.8&  6.8\\
 O  \sc viii    & 14.821&  14.825& 4.950e-03&    48.85&    13.98&  1377.9&   111.3&  6.5\\
 Fe \sc xvii    & 15.014&  15.015& 2.450e-03&   113.65&    17.53&  1703.0&   138.0&  6.7\\
 O  \sc viii    & 15.176&  15.180& 1.001e-04&   128.53&    18.47&  1339.7&   159.7&  6.5\\
 Fe \sc xvii    & 15.261&  15.266& 2.500e-03&    54.62&    13.85&  1446.7&   131.7&  6.7\\
 O  \sc viii    & 16.006&  16.010& 2.350e-03&   372.16&    25.82&  1548.3&   119.2&  6.5\\
 Fe \sc xvii    & 16.780&  16.775& 2.450e-03&    97.53&    19.92&  1330.7&   138.6&  6.7\\
 Fe \sc xvii    & 17.051&  17.050& 2.450e-03&   109.85&    20.63&  1075.7&   153.0&  6.7\\
 Fe \sc xvii    & 17.096&  17.095& 2.450e-03&   126.49&    21.40&  1075.7&   153.0&  6.7\\
 O  \sc vii     & 18.627&  18.627& 7.550e-03&    45.27&    24.42&  1500.0&   198.5&  6.3\\
 O  \sc viii    & 18.967&  18.972& 1.950e-03&  1958.10&    73.90&  1253.0&   223.0&  6.5\\
 O  \sc vii     & 21.602&  21.605& 2.649e-03&   222.86&    52.68&  1031.5&   193.3&  6.3\\
 O  \sc vii     & 21.804&  21.805& 5.000e-03&   130.96&    48.33&  1031.5&   193.3&  6.3\\
 O  \sc vii     & 22.098&  22.095& 2.600e-03&   204.20&    54.35&  1031.5&   193.3&  6.3\\
 N  \sc vii     & 24.779&  24.784& 5.051e-03&   124.05&    50.12&  1116.2&   311.6&  6.3\\       
 \enddata
 \tablenotetext{a}{Theoretical wavelengths of identification (from APED), in
   \AA. If the line is a multiplet, we give the wavelength of the
   stronger component.}
 \tablenotetext{b}{Measured wavelength, in \AA.}
 \tablenotetext{c}{Line flux is $10^{-6}$ times the tabulated value in $[\mathrm{phot\,cm^{-2}\,s^{-1}}]$.}
 \tablenotetext{d}{Continuum flux is $10^{-6}\times$ the tabulated value in $[\mathrm{phot\,cm^{-2}\,s^{-1}\,\AA^{-1}}]$.}
 \tablenotetext{e}{Logarithm of temperature [Kelvins] of maximum emissivity.}
 \tablenotetext{f}{Blended with high-$n$ Ne~{\sc x} Lyman series lines.}
 \end{deluxetable}
}

                                %\cbstart % refreply begin 4 [2001.05.09]
 {\small
 % ``modulation'' table
 \begin{deluxetable}{lp{2in}rrrr}
 \tablewidth{0pt}
 \tablecaption{Flux Modulation Line Groups\label{tab:fluxmod}}
 \tablehead{
   \colhead{Ion Group}&
   \colhead{Wavelengths\tablenotemark{a}}&
   \colhead{$f(Q)$\tablenotemark{b}}&
   \colhead{$\sigma$}&
   \colhead{$f(F)$\tablenotemark{b}}&
   \colhead{$\sigma$}
   }
 \startdata
Fe~{\sc xxv}     & 1.85& 0& 13& 32& 33\\
Fe~{\sc xxiv}    & 7.987, 10.620, 10.660, 11.176& 98& 15& 231& 25\\
Fe~{\sc xxi-xxii}& 11.743, 12.285&  109& 19& 146& 26\\
Fe~{\sc xvii}    & 15.015, 15.266, 16.775, 17.050, 17.095& 473& 62& 536& 89\\
O~{\sc viii}     & 14.825, 15.180, 16.010, 18.972& 2345& 119& 2834& 157\\
 \enddata
 \tablenotetext{a}{Theoretical wavelengths of identification (from APED), in
   \AA. If the line is a multiplet, we give the wavelength of the
   stronger component.}
 \tablenotetext{a}{Sum of line fluxes, $Q$ for Quiescent, and $F$ for Flare;
  $10^{-6}$ times the tabulated value gives $[\mathrm{phot\,cm^{-2}\,s^{-1}}]$.}
 \end{deluxetable}
}
                                %\cbend % refreply end 4 [2001.05.09]

%%%%%%%%%%%%%%%%%%%%%%%%%%%%%%%%%%%%%%%%%%%%%%%%%%%%%%%%%%%%%%%%%%%%
%t1  Ar=1.00 Fe=0.10 Mg=0.40 N=0.62 Ne=2.22 O=1.10 S=0.80 Si=0.46
%t2  Ar=1.00 Fe=0.15 Mg=0.50 N=0.89 Ne=2.57 O=1.06 S=0.47 Si=0.52
%tt  Ar=1.00 Fe=0.15 Mg=0.48 N=0.65 Ne=2.35 O=1.10 S=0.65 Si=0.45

 {\small
 % Abundance table
 \begin{deluxetable}{ccccc}
 \tablewidth{0pt}
 \tablecaption{Abundance Determinations\label{tab:abund}}
 \tablehead{
   \colhead{Element}&
   \colhead{FIP\tablenotemark{a}}&
   \colhead{Quiescent\tablenotemark{b}}&
   \colhead{Flare\tablenotemark{b}}&
   \colhead{Total\tablenotemark{b}}
   }
 \startdata
 Mg &  7.65  & 0.40  & 0.50 & 0.50\\
 Fe &  7.87  & 0.10  & 0.15 & 0.15\\
 Si &  8.15  & 0.45  & 0.50 & 0.45\\
 S  & 10.36  & 0.80  & 0.45 & 0.65\\
 O  & 13.62  & 1.10  & 1.05 & 1.10\\
 N  & 14.53  & 0.6:  & 0.9: & 0.6:\\
 Ar & 15.76  & 1:    & 1:   & 1:  \\
 Ne & 21.56 & 2.20  & 2.60 & 2.35\\
 \enddata
 \tablenotetext{a}{First Ionization Potential, in eV}
 \tablenotetext{b}{Ratio of abundances to cosmic values of 
   \citet{Anders89}; uncertainties
   are about 20\% (50\% for values marked with ``:'').}
 \end{deluxetable}
 }

\end{document}